\documentstyle[aps,epsf,overcite]{revtex}
\begin{document}
\newcommand{\beq}{\begin{equation}}
\newcommand{\eeq}{\end{equation}}
\newcommand\bea{\begin{eqnarray}}
\newcommand\eea{\end{eqnarray}}
\def\a{\alpha}
\def\b{\beta}
\def\g{\gamma}
\def\G{\Gamma}
\def\h{\hat}
\def\ch{\chi}
\def\ep{\epsilon}
\def\ph{\phi}
\def\s{\sigma}
\def\th{\theta}
\title{Full Optimization of Linear Parameters\\
of a United Residue Protein Potential }
\author{Julian Lee, Kibeom Park$^\dagger$ and Jooyoung Lee$^*$}
\address{School of Computational Sciences\\
Korea Institute for Advanced Study\\
Seoul 130-012, Korea}
\maketitle
\begin{abstract}
 We apply the general protocol of parameter optimization
 (Lee, J. et al. {\it Phys. Chem. B} {\bf 2001}, {\sl 105}, 7291)
 to the UNRES potential.
 In contrast to the earlier works where only the relative weights of
 various interaction terms were optimized,
 we optimize all linear parameters of the potential.
 The method exploits the high efficiency of the
 conformal space annealing method in finding distinct low
energy conformations. For a given training set of proteins, the
parameters are modified to make the native-like conformations
energetically more favorable than the non-native ones. Linear
approximation is used to estimate the energy change due to the
parameter modification. The parameter change is followed by local
energy reminimization and new conformational searches to find the
energies of native-like and non-native local minima of the energy
function with new parameters. These steps are repeated until the
potential predicts a native-like conformation as one of the low
energy conformations for each protein in the training set. We
consider a training set of crambin (PDB ID 1ejg), 1fsd, and the
10-55 residue fragment of staphylococcal protein A (PDB ID 1bdd).
As the first check for the feasibility of our protocol, we optimize
the parameters separately for these proteins and find an optimal
set of parameters for each of them. Next we apply the method
simultaneously to these three proteins. By refining all linear
parameters, we obtain an optimal set of parameters from which the
native-like conformations of the all three proteins are retrieved
as the global minima, {\it  without introducing additional
multi-body energy terms}.
\end{abstract}
$^*$ To whom the correspondence should be addressed. E-mail: jlee@kias.re.kr.\\
$^\dagger$ Present address: Electronics and Telecommunications Research Institute, Daejon, 305-350, Korea.
\section{Introduction}
\setcounter{equation}{0}

The prediction of the three dimensional structure of a protein
solely from its amino acid sequence is one of the most challenging
problems in computational sciences today. Popular approaches to
this problem have been comparative modeling and fold recognition,
which can be classified as knowledge-based
methods\cite{know1,know2,know3,know4}. These methods use
statistical relationship between the sequences and the
three-dimensional structures of the proteins in the Protein Data
Bank (PDB) in order to predict the unknown structure of a protein
sequence, without deep understanding of the protein folding.

 On the other hand, the {\it ab initio} method\cite{ab1,ab2,corn1,corn2,corn3,corn4,corn6},
 which is also called the energy-based or physics-based method, is based on
 the thermodynamic hypothesis which postulates that proteins adopt native structures that minimize their
 free energies\cite{anfin}.
 Since it attempts to understand the fundamental principles of the protein folding itself, the success of
 this method will lead not only to the
 successful structure prediction, but also to the clarification of the protein folding mechanism.

 However, there have been two major obstacles to the successful application of energy-based methods to the protein folding problem.
 First, the energy landscape of a protein is riddled with an astronomical number of local minima, making it
 difficult to search. Secondly, there are inherent inaccuracies in potential energy functions which attempt to describe the energetics of proteins.
 The first problem has been largely alleviated to some extent by recent developments of efficient search algorithms
 such as the conformational
 space annealing (CSA) method\cite{csa1,csa2,csa3}. The second problem is the one which is addressed in this paper.
The accuracy of a given potential energy function can be improved
by modifying its functional form as well as its parameters. In
this work, we will refine the parameters of the potential energy
without changing the functional form.

  Physics-based potentials are generally parameterized from quantum mechanical calculations and experimental
  data on model systems\cite{jlee}. However, such calculations and data do not determine the parameters with
  perfect accuracy.
  The residual errors in potential energy functions may have significant effects on simulations of macromolecules such as
  proteins
  where the total energy is the sum of a large number of interaction terms. Moreover, these terms are known to cancel each
   other to a high degree, making their systematic errors even more significant. Thus it is crucial to refine
   the parameters of a potential energy function before it is applied to the protein folding problem.

 In fact, an iterative procedure which systematically refines the parameters of a given
 potential energy function was presented by Lee et al.\cite{jlee}. Since the CSA method can
efficiently sample a wide range of the conformational space of a protein, the
strategy is to apply this method to the
 proteins with known structures in order to refine the potential.
 We refine the parameters so that native-like conformations of
 these proteins have lower energies than non-native ones. The
 set of the proteins used for the parameter refinement is called
 the training set. It would be desirable to include many proteins
 in the training set,
 which belong to representative structural classes of
 proteins. However, it is quite a non-trivial issue to check whether this procedure itself is feasible, even for a small number of proteins in the training set.

 The 10-55 fragment of staphylococcal protein A (1bdd) was used by Lee et al.\cite{jlee}
 to refine a coarse grained potential called the UNRES potential\cite{liwo3,liwo4,liwo5},
 where each residue is approximated by two interaction sites.
 This potential was successfully applied for predicting the unknown structures
 of proteins in CASP3\cite{corn1,corn4,casp3}, and its basic version consists of
seven interaction terms. Lee et al.\cite{jlee} optimized six relative weights
of these interaction terms. Since only an $\alpha$
protein was used for the refinement, the resulting potential was
suitable for $\a$ proteins.

 On the other hand, Pillardy et al.\cite{pill} used three training sets
consisting of one protein 1pou, one protein 1tpm, and two proteins
betanova and 1bdd, to optimize the potentials for predicting
$\a,\b$, and $\a / \b$ proteins, respectively\cite{foot1}. This
potential was extended to include six additional multi-body terms,
which increased the number of relative weights to be optimized
from six to twelve\cite{foot2}. Therefore in these works, the
functional form of the UNRES potential as well as a part of the
parameters was modified, which are the twelve relative weights.
The introduction of the six additional multi-body terms were
necessary in order to incorporate proteins with $\beta$ strands.

However, one should note that each of the seven interaction terms in the original version of the UNRES potential has its own parameters
in it. Therefore it is a natural question to ask whether one can
optimize the potential energy function for the proteins with $\b$-strands by refining these
parameters, without introducing additional multi-body energy terms. Of course, it might not be possible to optimize the potential for arbitrarily many proteins without introducing additional interaction terms. However, it is important to optimize the parameters as much as possible before introducing functional modifications, since this will give us better insights on the limitations of a given potential and the types of additional interaction terms necessary for improvement.

Indeed we observe that it is possible to refine the UNRES
potential with three proteins 1bdd, 1fsd, and crambin (1ejg) {\it
without introducing additional multi-body terms}, where proteins 1fsd and
1ejg contain $\beta$ strands. First, the parameters are optimized
separately for these proteins, and an optimal parameter set for
each of them is obtained. The potentials with optimized parameter
sets yield global minimum energy conformations (GMECs) with
root mean square deviations (RMSDs) of
 $1.7$, $2.5$, and $2.6$ \AA \ ffrom the native structures, for 1bdd, 1fsd, and 1ejg respectively.
Finally the parameters are refined for the
training set consisting of these three proteins, and a
parameter set is obtained which correctly describe the energetics of these proteins
simultaneously. The potential with the optimized parameter set yields the GMECs with RMSDs of $1.8$, $2.5$, and $2.6$ \AA, respectively.


\section{Methods}

\subsection{General Protocol}
A brief description of our procedure is as follows. In order to
check the performance of a potential energy function for a given
set of parameters, one has to sample native-like conformations as
well as non-native ones. Non-native conformations can be obtained
by an unrestricted conformational search which we call global CSA. Native-like
conformations are obtained by a restricted search which we call
local CSA.
   In the local CSA, only the conformations whose RMSDs from the native structure are below a preset cutoff value, are sampled.

   Since a potential can be considered to describe the nature correctly if native-like structures have lower energies than the non-native ones, the optimization  criterion is given in terms of the energy gap, which is the difference between the lowest energy of the native-like conformations and that of the non-native ones. We define the energy gap to be negative when the lowest energy of the native conformations is lower than that of non-native ones.
We modify the parameters so that the energy gaps of the proteins in the training set decrease.
The changes of energy gaps are estimated by the linear approximation
of the potential in terms of parameters (See section D.).
Since the positions of the local energy minima are shifted due to the parameter modification,
it is necessary to reminimize their energies with the new parameters. We also search the conformational space
with the newly obtained parameters to find new low-lying local energy minima.
Together with the energy-reminimized conformations, these constitute a structural database which will be used for subsequent refinement of the parameters. We iterate these steps until the energy gaps become all negative for proteins in the training set. The detailed explanation for each step of our procedure is given below.

\subsection{Potential Energy Function}
We use the UNRES force field\cite{liwo3,liwo4,liwo5}, where a
polypeptide chain is represented by a sequence of $\alpha$-carbon
($\rm C^\alpha$) atoms linked by virtual bonds with attached
united side chains (SC) and united peptide groups (p) located in
the middle between the consecutive $\rm C^\alpha$'s. All the
virtual bond lengths are fixed; the $\rm C^\alpha$-$\rm C^\alpha$
distance is taken as $3.8$ \AA, and $\rm C^\alpha$-$\rm SC$
distances are given for each amino acid type. The energy of the
chain is given by
\begin{eqnarray}
E&=& \sum_{i<j} U_{\rm SCSC}(i,j) + w_{\rm SCp} \sum_{i\neq j}
U_{\rm SCp}(i,j)
     + w_{\rm pp} \sum_{i<j-1} U_{\rm pp}(i,j) +
       w_{\rm b} \sum_i U_{\rm b}(i) \nonumber \\
 & & + w_{\rm tor} \sum_i U_{\rm tor}(i)+
       w_{\rm rot} \sum_i U_{\rm rot}(i)
      +w_{\rm dis} U_{\rm dis} + w^{(4)}_{\rm el-loc} \sum_{i<j} U^{(4)}_{\rm el-loc}(i,j),\label{pot}
\end{eqnarray}
where $w$'s are the relative weights which were refined in the
earlier works\cite{jlee,pill,liwo7}. As described in detail in the
appendix, $U_{\rm SCSC},U_{\rm SCp}, U_{\rm pp},U_{\rm tor}$, and
$U^{(4)}_{\rm el-loc}$ can be further decomposed into linear
combinations of smaller parts, whose coefficients are refined in
this work. Therefore we may fix the values of $w_{\rm SCp}, w_{\rm
pp}$, $w_{\rm tor}$, and $w^{(4)}_{\rm el-loc}$ without loss of
generality. We set them to unity for simplicity. Here, $U_{\rm
SCSC}(i,j)$ represents the mean free energy of the hydrophobic
(hydrophilic) interaction between the side chains of residues $i$
and $j$, which is expressed by Lennard-Jones potential, $U_{\rm
SCp}(i,j)$ corresponds to the excluded-volume interaction between
the side chain of residue $i$ and the peptide group of residue
$j$, and the potential $U_{\rm pp}(i,j)$  accounts for the
electrostatic interaction between the peptide groups of residues
$i$ and $j$. The terms $U_{\rm tor}(i)$, $U_{\rm b}(i)$, and
$U_{\rm rot}(i)$, denote the short-range interactions,
corresponding to the energies of virtual dihedral angle torsions,
virtual angle bending, and side chain rotamers, respectively.
$U_{\rm dis}$ denotes the energy term which forces two cystein
residues to form a disulfide bridge.  Finally, the four-body
interaction term $U^{(4)}_{\rm el-loc}$ results from the cumulant
expansion of the restricted free energy  of the polypeptide chain.
In contrast to the earlier works\cite{pill,liwo7} where additional
multi-body terms were introduced\cite{liwo6}, $U^{(4)}_{\rm
el-loc}$ is the only multi-body term used in this work. The
detailed forms of these terms are given in the appendix. As
discussed there, the total number of linear parameters which we
adjust is 709. The functional form Eq.(\ref{pot}), as well as the
initial parameter set we use, is the one used in the CASP3
exercise\cite{corn1,casp3}.

\subsection{Global and Local CSA}
In the protein folding problem, the energy surface contains an
astronomical number of local energy minima. The larger a protein is, the
more likely it is that there exist many local energy minima which
correspond to very different structures. In general,
it is not sufficient to consider only the lowest
energy conformation as a possible candidate for the native
structure. Since the force field parameters contain inevitable
errors, one should take account of many distinct low energy
conformations. Therefore it is necessary to search the whole conformational space.

It has been shown that this multiple minima problem can be
overcome by an efficient search algorithm such as the CSA method.
In this work, extensive conformational searches are carried out by
global and local CSA methods\cite{csa1,csa2,csa3}. The  CSA method
can be considered as a genetic algorithm that enforces a broad
sampling in its early stages and gradually allows the
conformational search to be focused into narrow conformational
space in its later stages. As a consequence, many low-energy local
minima including the GMEC of the benchmark protein can be
identified for a given parameter set. Unless the parameters are
properly optimized, these conformations can be quite different
from the native structure. Therefore, in this case, we may
consider the global CSA as the sampling of the non-native
conformations. On the other hand the native-like conformations are
sampled by the local CSA search\cite{jlee}. The local CSA is the
restricted search where only the conformations whose $\rm
C^\alpha$ RMSD values are within a fixed cutoff, $R_c$, of the
native conformation, are sampled. Also, in order to find these
native-like structures, the initial conformations are prepared
with the native backbone coordinates, whose energy is subsequently
minimized\cite{foot3}. The value of $R_c$ should be large enough
to sample representative native-like conformations and at the same
time small enough to eliminate non-native conformations.

\subsection{Linear Approximation and Parameter Refinement}
\begin{figure}
\epsfxsize=15cm \epsfbox{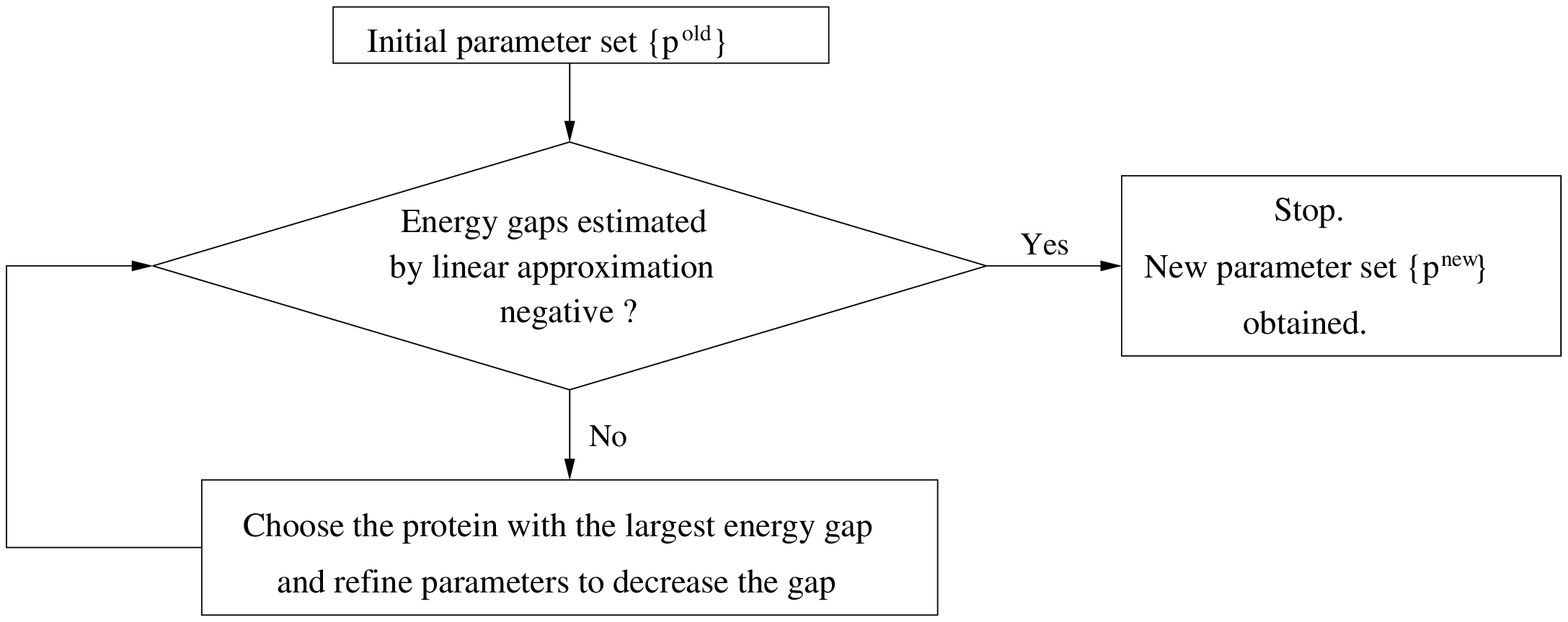}
\vskip 0.5cm
\caption{The flow
chart for the part of the algorithm where the parameters are
refined using linear approximation for the energy gaps.}
\label{flow1}
\end{figure}

  Once the energies of the non-native and native-like conformations for all proteins in the training set are obtained, the parameters are modified as follows.
  We select the protein with the largest energy gap,
  and change the parameters so that this
  energy gap decreases.
   The parameters are changed by small amounts at each step, so the energy with the new parameters can be
   estimated by
  the linear approximation:

\begin{equation}
\label{Erem}
E^{\rm new} \approx E^{\rm old}+ \sum_i (p^{\rm new}_i-{p}^{\rm old}_i)
         \frac{\partial E^{\rm old}}{\partial p_i},
\end{equation}
where the $p^{\rm old}_i$'s and $p^{\rm new}_i$'s represent the parameters before and after
modification, respectively.
The parameter dependence of the position of the local minimum can
be neglected in the linear approximation, since the derivative in
the conformational space vanishes at a local minimum\cite{jlee}.
In general the derivative $\frac{\partial E}{\partial p_i}$ is a
function of the parameters, but for linear parameters it is just a
constant independent of parameters. In this work, we adjust only
the linear parameters for simplicity, the total number of them
being 709 for the UNRES potential. The details can be found in the
appendix. Therefore the energy function can be written as:
\begin{equation}
E=\sum_i p_i e_i.
\end{equation}
where $e_i$'s are the coefficients independent of $p_i$.
The change of the energy gap is estimated as:
\begin{eqnarray}
\Delta E_{\rm gap} &=& E_{\rm gap}(\{ p^{\rm new}_j \})- E_{\rm
gap}(\{ p^{\rm old}_j \})\nonumber\\
&=&(E^{\rm (lowest ~N)}(\{ p^{\rm new}_j \})-E^{\rm (lowest ~NN)}(\{ p^{\rm new}_j \}))-(E^{(\rm lowest ~N)}(\{ p^{\rm old}_j \})-E^{\rm (lowest ~NN)}(\{ p^{\rm old}_j \}))\nonumber\\
&=& \sum_i [e_i({\rm lowest ~N})-[e_i({\rm lowest
~NN})](p^{\rm new}_i-p_i^{\rm old}). \label{linapp}
\end{eqnarray}
where $E$ and $e$ are evaluated for the
lowest energy native-like (N) and non-native (NN) conformations.
We fix the magnitude of the parameter change $\delta p_i \equiv p_i^{\rm new}-p_i^{\rm old}$ to be a certain fraction $a$ of $p^{\rm old}_i$.
We use $a=0.01$ in this study.
The sign of $\delta p_i$ is chosen to decrease the energy gap,
\begin{equation}
\delta p_i = -a p_i^{\rm old} sign[e_i({\rm lowest ~N})-e_i({\rm lowest
~NN})]. \label{lpch}
\end{equation}
We repeat this procedure of selecting the protein with the largest
energy gap and modifying the parameters, until the energy gaps
estimated by Eq.(\ref{linapp}) become all negative for proteins in
the training set. The flow chart for this part of the algorithm is
shown in Fig.~\ref{flow1}.

\subsection{Reminimization and new conformational search}
 Since the procedure of the previous section was based on
 the linear approximation Eq.(\ref{linapp}) and the number of
 conformations in the structural database is limited,
 we now have to evaluate the true energy gap
 using the newly obtained parameters. The breakdown of the parameter refinement may come
  from two sources.
  First,
  the conformations corresponding to the local minima of
  the potential
  for the original set of parameters are no longer necessarily so for the new
  parameter set. For this reason, we reminimize the energy of these
  conformations with the new parameters.
Secondly, the local minima obtained using CSA method with the
original parameter set are only a tiny fraction of the whole set
of local minima. After the change of the parameters, some of the
local minima which were not considered due to their relatively
high energies, can now have low energies for the new parameter
set. It is even possible that entirely new low-energy local minima
appear. Therefore these new minima are taken into account
by performing
 subsequent CSA searches with the newly obtained parameter set.

\subsection{Update of the structural database and iterative refinement of parameters}
\begin{figure}
\epsfysize=10cm \epsfbox{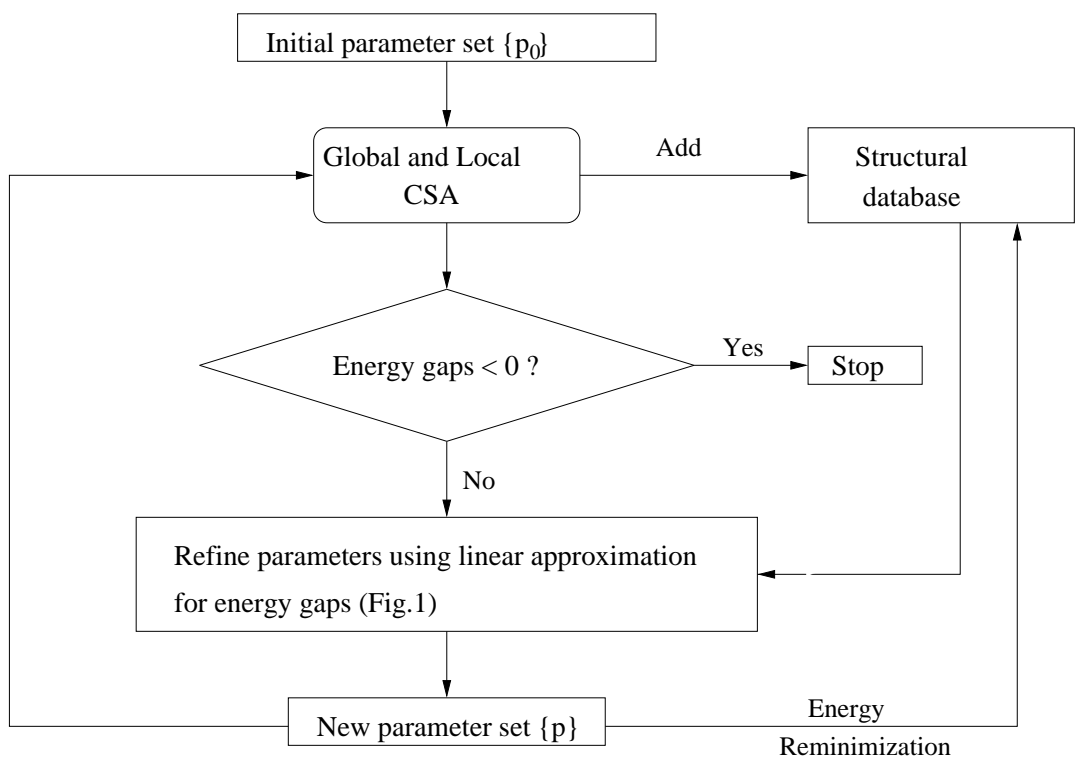}
\vskip 0.5cm \caption{The flow
chart for the whole protocol of the iterative parameter
refinement. The parameter refinement step using the linear
approximation for the energy gap corresponds to the flow chart of
Fig.~1.} \label{flow2}
\end{figure}

\begin{figure}
\epsfysize=21cm
\epsfbox{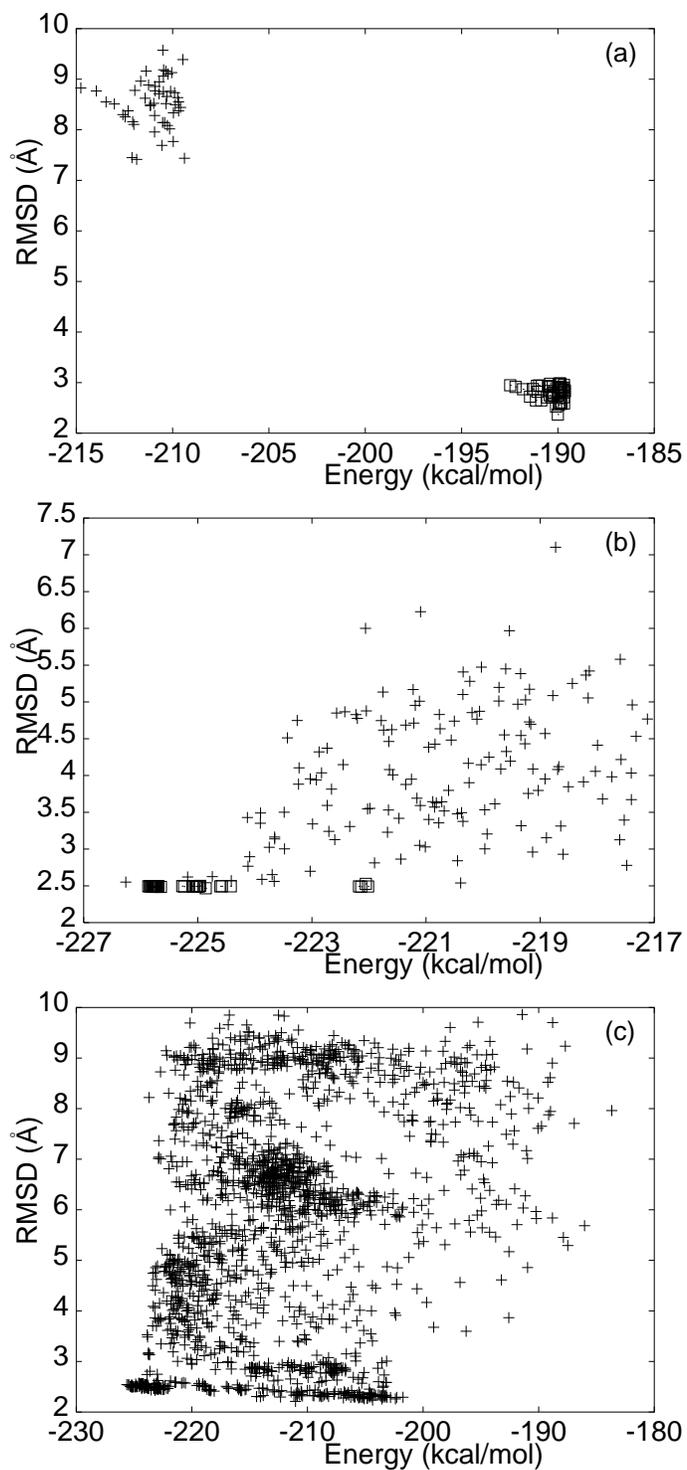}
\caption{Plots
of the UNRES energy and ${\rm C^\alpha}$ RMSD (from the native
structure) for conformations of 1ejg obtained from global and local
CSA searches, and the ones in the structural database. (a) The
result with the initial parameter set. We observe that the GMEC is of
RMSD $8.8$ \AA , and there are native-like conformations
with RMSDs less than $2.5$ \AA. (b) The final result after 13 iterations. We observe
that the GMEC is of RMSD $2.6$ \AA.
(c) The plot of the conformations in
the structural bank whose energies are reminimized with the
optimized parameter set. } \label{crs}
\end{figure}

 The low-lying local energy minima found in the new conformational searches are added to the
  energy-reminimized conformations to form a structural database of local
  energy minima. The conformations in the database are used to obtain the energy gaps, and if
  their values are not satisfactory, these conformations are used for the new
  round of parameter refinement. As the procedure of [CSA $\rightarrow$
  parameter refinement $\rightarrow$ energy reminimization] is repeated,
  the number of conformations in the structural database increases.
  As an example, the energy-RMSD plot of an energy-reminimized structural database for  the protein 1ejg is shown in Fig.~\ref{crs}(c).
  The flow chart of the whole procedure is
  illustrated in Fig.~\ref{flow2}. This  iterative procedure is continued until the
  energy gaps become negative for all proteins.

  \subsection{Choice of RMSD cutoff}

   It is important to choose the RMSD cutoff judiciously for each protein
   in the training set, in order to carry out the whole procedure efficiently. This cutoff is the criterion for
   distinguishing the native-like and non-native conformations.
   The cutoff is necessary in two places in the procedure. First, it is used in
   the local CSA where conformations with RMSDs below a preset cutoff value are sampled, and secondly, in
   the parameter refinement step where the conformations in the structural database
   are divided into native-like and non-native families. In general, these two cutoff values
   can be different from each other. In addition, a separate cutoff value can be used for each iteration. In this work we check the
   distribution of RMSD {\it  vs.} energy of the conformations by visual inspection, and cluster them into native-like and
   non-native families to determine the appropriate values of RMSD cutoff for each protein.
   The values of RMSD cutoff used are given in
   table \ref{rmstable1} and \ref{rmstable2}.

\section{Results}
  We consider a training set consisting of three proteins.
  They are the 10-55 fragment of the B-domain of staphylococcal protein
  A (1bdd), 1fsd,  and crambin (1ejg), which are 46, 28, 46 residues long respectively.
  We first refine parameters for
these proteins separately, to check whether our
protocol for the iterative parameter refinement is feasible.

\subsection{Separate parameter refinement for each protein}

This is the simplest case of having one protein in the training
set.  The first example is the 1bdd, which was the target protein
of the previous study of the weight optimization\cite{jlee}. In
that work, a negative energy gap was found after 6 iterations, and
the GMEC which has a $2.2$ \AA\  RMSD deviation from native
structure was obtained. We start with the same initial parameters,
which were used in CASP3\cite{jlee,corn1,casp3}.

\begin{figure}
\epsfysize=20cm \epsfbox{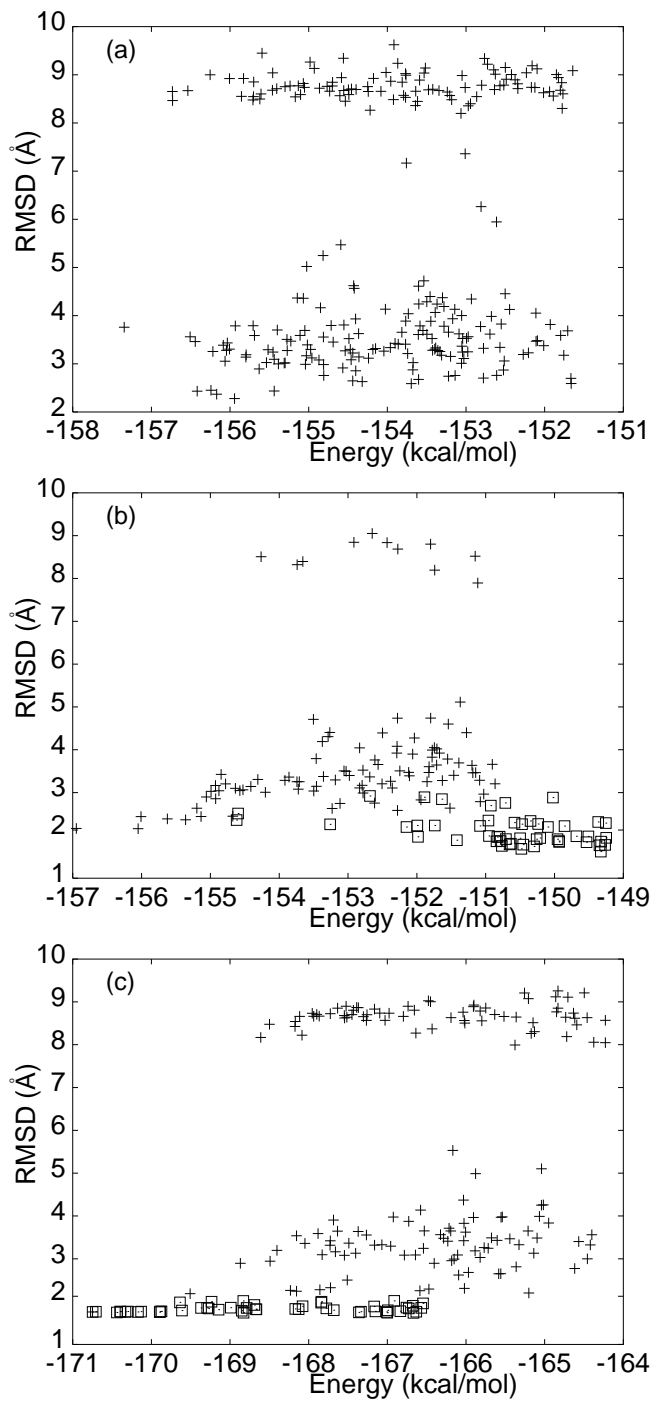} \vskip 0.5cm \caption{Plots of
the UNRES energy and ${\rm C^\alpha}$ RMSD (from the native
structure) for conformations of 1bdd obtained from CSA searches.
(a) The result with the initial parameter set. We observe that the
GMEC is of RMSD $3.8$ \AA , and there are native-like
conformations with RMSDs less than $2.4$ \AA. (b) The result after
3 iterations. In all the figures in this paper, the plus signs
denote the conformations from the global CSA, and the squares
denote those from the local CSA. We observe that the GMEC is of
RMSD $2.2$ \AA, and there are native-like conformations with RMSDs
less than $2.0$ \AA. (c) The final result after 11 iterations. We
observe that the GMEC is of RMSD $1.7$ \AA.} \label{pas}
\end{figure}

\begin{figure}
\epsfysize=21cm
\epsfbox{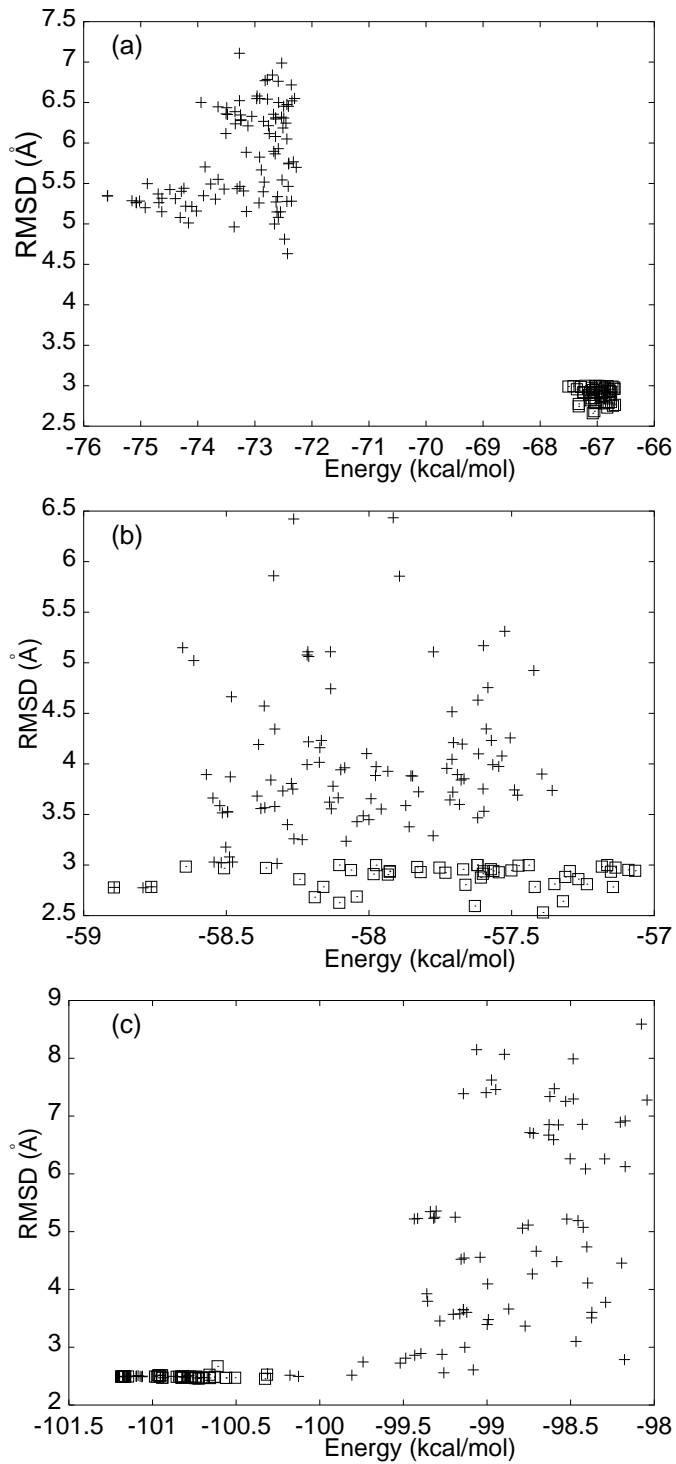}
\caption{Plots of the UNRES energy
and ${\rm C^\alpha}$ RMSD (from the native structure) for
conformations of 1fsd obtained from global and local CSA searches.
(a) The result with the initial parameter set. We observe that the
GMEC is of RMSD $5.4$ \AA , and there are native-like
conformations with RMSDs less than $2.8$ \AA. (b) The result after 3
iterations. We observe that the GMEC is of RMSD $2.8$ \AA, and there are
native-like conformations with RMSDs less than $2.6$ \AA. (c) The final result
after 12 iterations. We observe that the GMEC is of RMSD $2.5$ \AA.
} \label{fss}
\end{figure}

In the CSA sampling with the original parameter set, the GMEC was of
$3.8$ \AA\ RMSD from the native structure\cite{corn2}, as shown in
Fig.~\ref{pas}(a). We set $R_c=3.0$ \AA\ , adjust the parameters
according to it, and proceed to the next iteration. A negative energy gap
is found after 3 iterations of parameter refinement with $R_c=2.2$
\AA\ and the global CSA search yields the global minimum at $2.2$
\AA. Furthermore, from the local CSA run we find conformations
with RMSDs lower than $2.0$ \AA\ (Fig.~\ref{pas}(b)). Therefore
we repeat the procedure with lower values of $R_c$. After 11
iterations the GMEC with RMSD$=1.7$ \AA\  is obtained
(Fig.~\ref{pas}(c)). We have further proceeded with an even lower
value of $R_c$. However, the energy gap does not improve after the
11-th iteration. Therefore we take the result from the 11-th
iteration as the final optimized parameter set for this protein.
The results are shown in Fig.~\ref{pas}(c). Similar procedures are
repeated for 1fsd and 1ejg to obtain optimized parameter sets
which yield GMECs with RMSD values $2.5$ \AA \ and $2.6$ \AA \ respectively. Details are shown in tables \ref{rmstable1},
\ref{rmstable2} and figures \ref{pas}, \ref{fss}, \ref{crs}.

\subsection{Simultaneous parameter refinement for three proteins}

Again, the initial parameter set is the one used in
CASP3\cite{corn1,casp3}. Since a large number of conformations
were already accumulated in the structural databases during the
separate parameter optimizations for three proteins, we use them
to start the iterative procedure of simultaneous parameter
refinement for the three proteins.

\begin{figure}
\epsfysize=21cm
\epsfbox{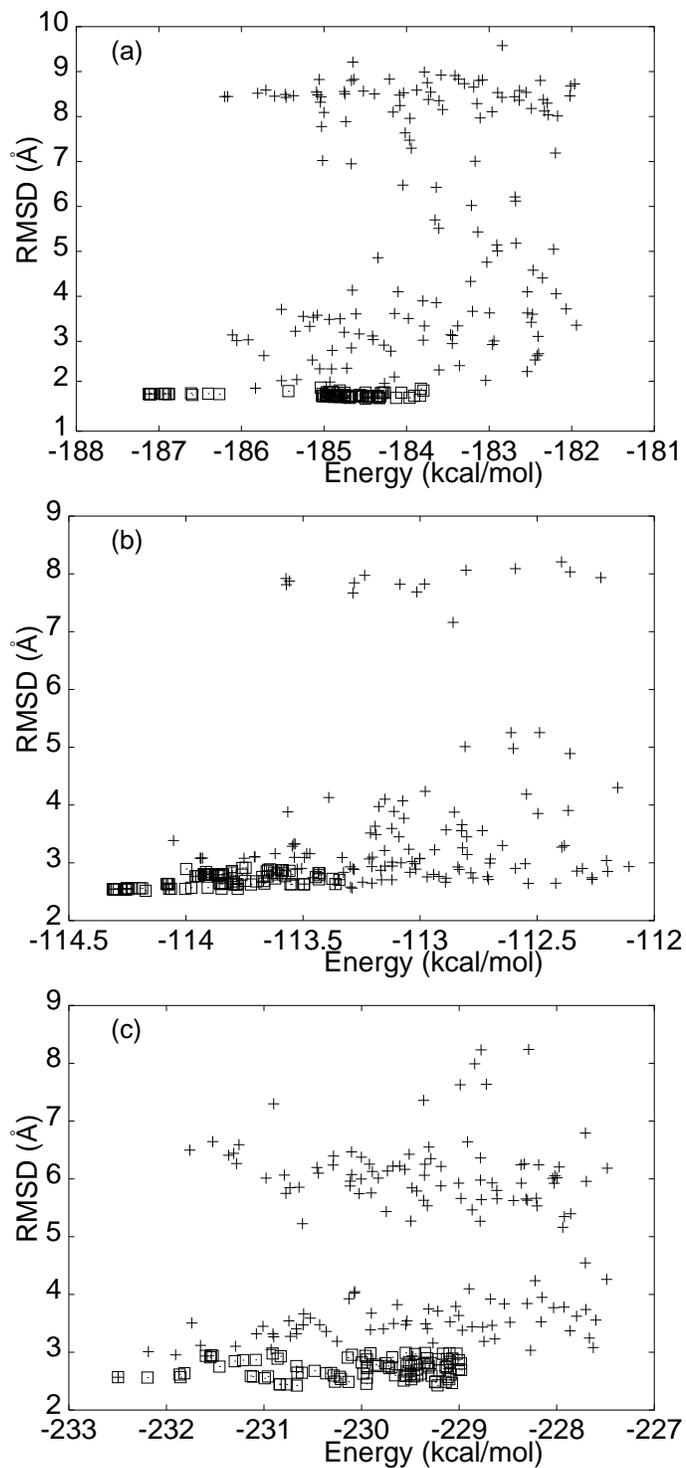}
\caption{Plots of the UNRES energy and
${\rm C^\alpha}$ RMSD (from the native structure) for
conformations of the three proteins obtained from global and local
CSA searches with the parameter set which is optimized for these
proteins simultaneously. (a) The plot for 1bdd. We observe that the
GMEC is of RMSD $1.8$ \AA. (b) The plot for 1fsd. The
GMEC is of RMSD $2.5$ \AA. (c) The plot for 1ejg. The
GMEC is of RMSD $2.6$ \AA.} \label{aa6}
\end{figure}

\begin{figure}
\epsfysize=21cm \epsfbox{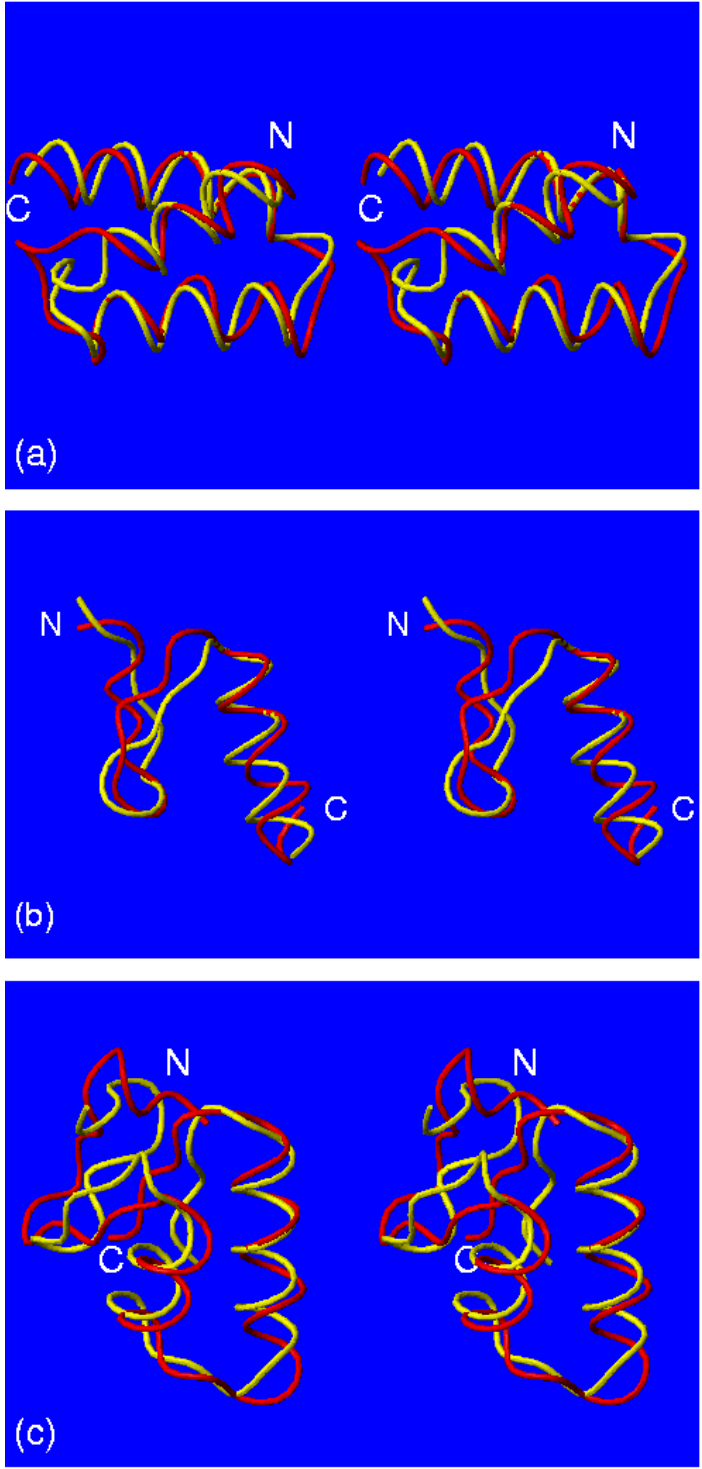}
\caption{(a) The ${\rm C^\alpha}$ trace of 1bdd. The native
structure is shown in red and the GMEC found with the optimized
parameters is shown in yellow. The RMSD is $1.8$ \AA. (b) The
${\rm C^\alpha}$ trace of 1fsd. The native structure is shown in
red and the GMEC found with the optimized parameters is shown in
yellow. The RMSD is $2.5$ \AA. (c) The ${\rm C^\alpha}$ trace of
1ejg. The native structure is shown in red and the GMEC found with
the optimized parameters is shown in yellow. The RMSD is $2.6$
\AA. The figures were prepared with the program
MOLMOL\cite{molmol}.}\label{pro3}
\end{figure}

\begin{figure}
\epsfxsize=16cm
\epsfbox{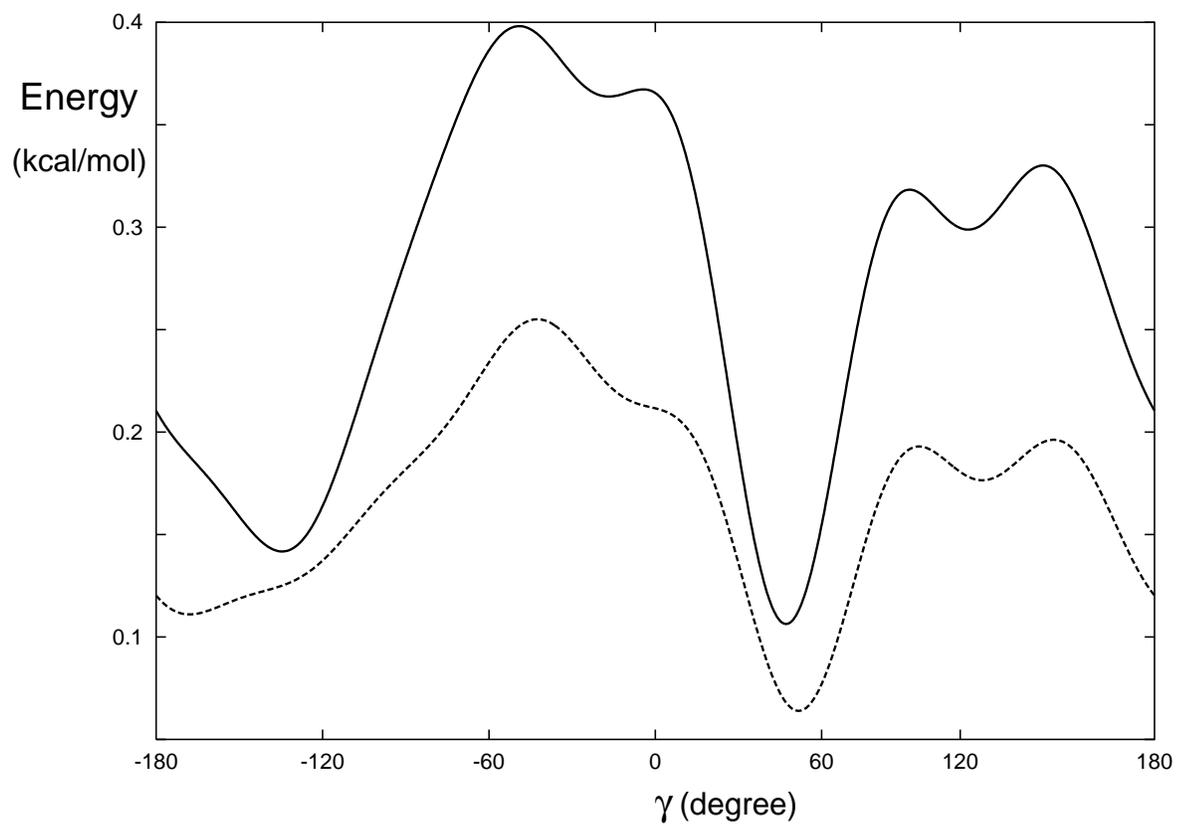}
\caption{ The torsional potential between the residues which are neither glycine nor proline, as a function of the torsional angle $\g$.  The solid (dashed) line is obtained with the optimized (original) parameters. }
 \label{etor}
\end{figure}

By choosing appropriate values of RMSD cutoff for each iteration,
we obtain an optimized parameter set after 6 iterations, yielding
the GMECs with RMSD values of $1.8$ \AA, $2.5$ \AA, and $2.6$ \AA\ for
1bdd, 1fsd, and 1ejg, respectively. The energies and RMSDs of the
conformations obtained with the optimized parameters are plotted
in Fig.~\ref{aa6}, and the  $\rm C^\alpha$ trace of the GMEC conformations
are shown in Fig.~\ref{pro3} along with
the native conformations. The numerical values of the optimized
parameters are provided in the Supporting Information. As an example of the changes of interaction terms due to the parameter optimization, the torsional
energies between residues which are neither glycine nor proline, with the optimized and original parameter set, are plotted in
Fig.~\ref{etor}.


\section{Jackknife test}

  It should be noted that the purpose of the present work is
  not to provide a potential which is transferable to all proteins, but to develop a methodology for optimizing  potential parameters of {\it a given potential}.
  Applying this method to develop a transferable potential is out
  of the scope of this paper, and a much larger training set and even additional interaction terms might be necessary in order to achieve it.
   In fact, it is quite nontrivial to check whether such a procedure
  is possible at all. However, we performed conformational searches for proteins not included in the training set, which is called a Jackknife test, and find some interesting features.

\begin{figure}
\epsfysize=11cm \epsfbox{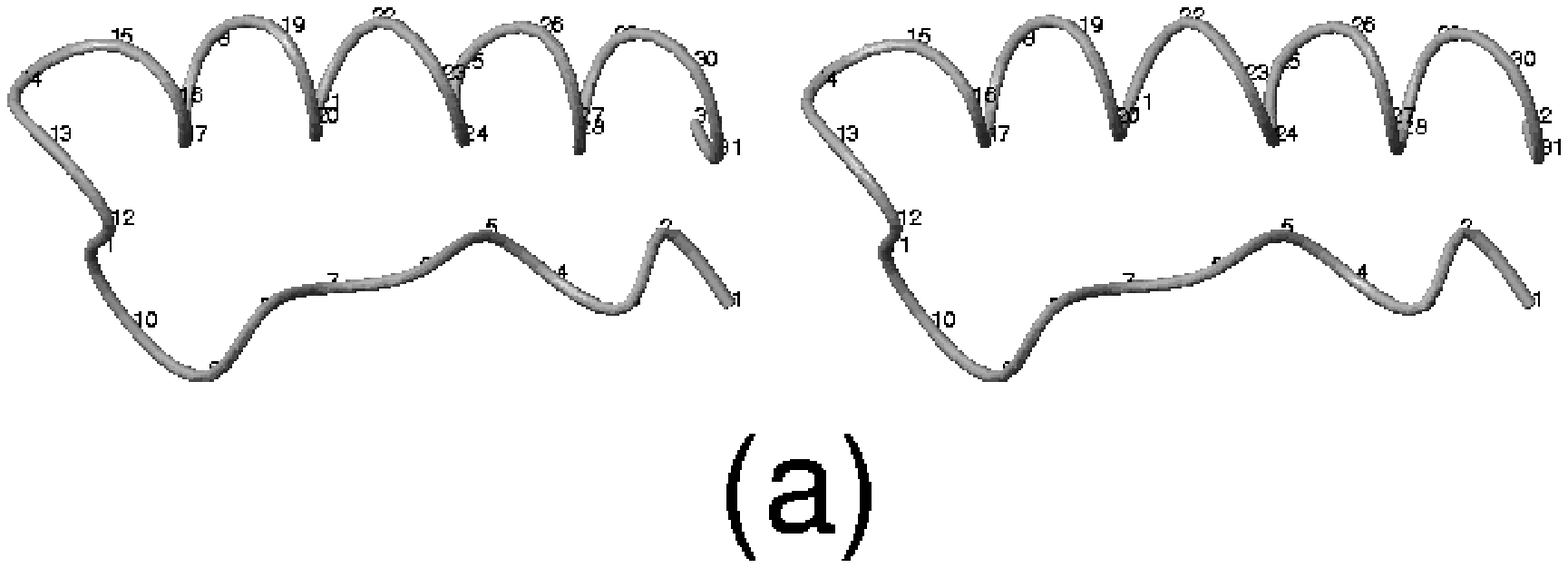} \epsfysize=11cm
\epsfbox{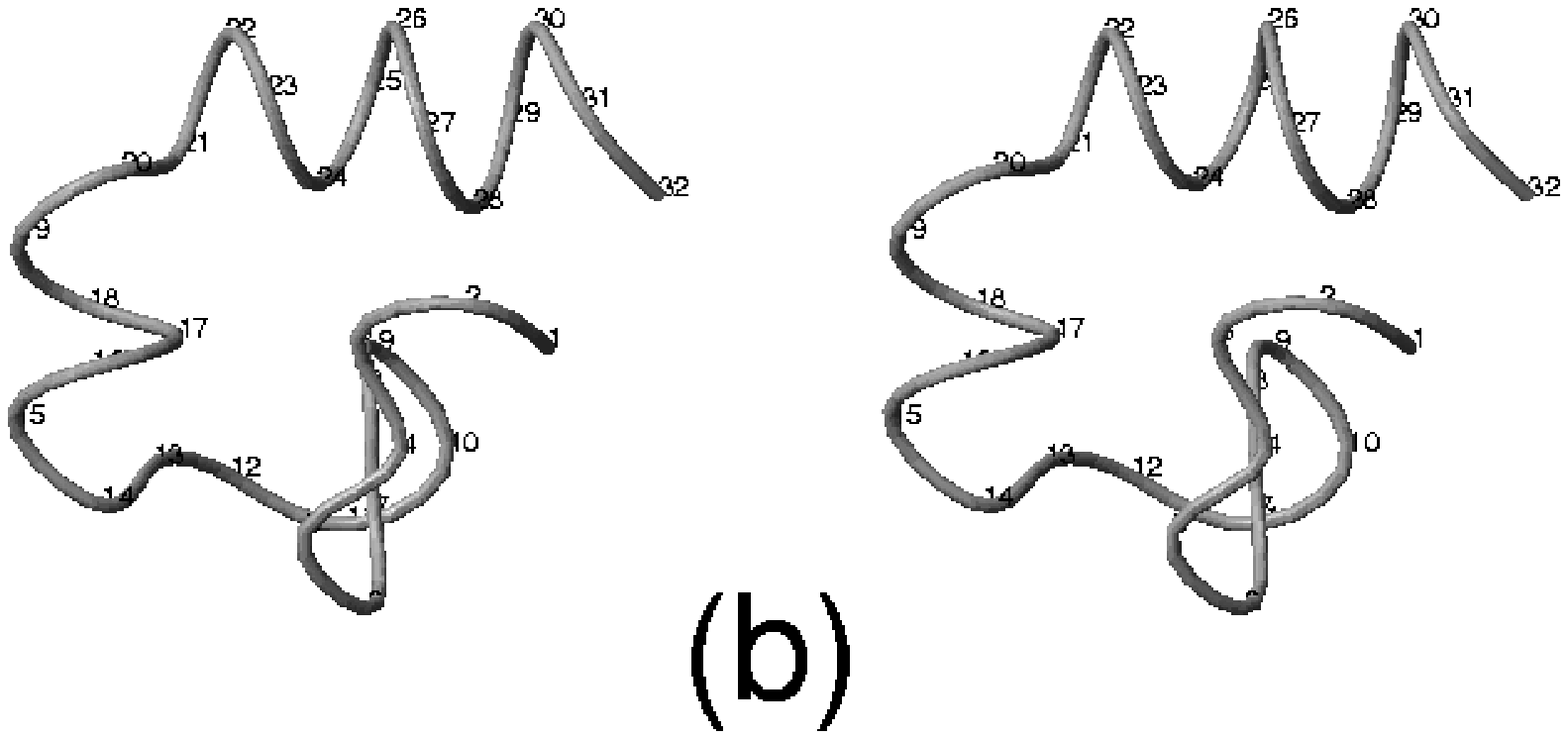}
\newpage
\epsfysize=11cm \epsfbox{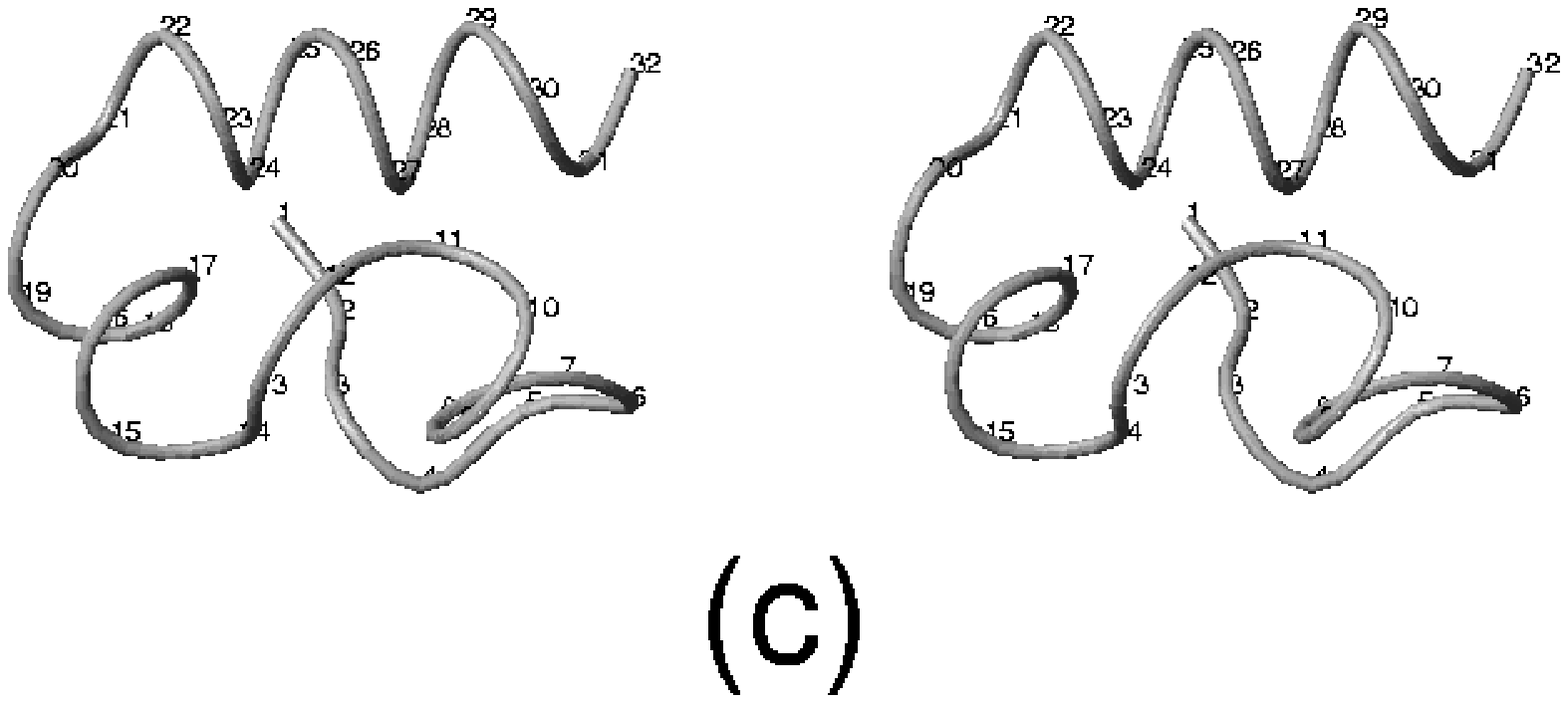} \epsfysize=11cm
\epsfbox{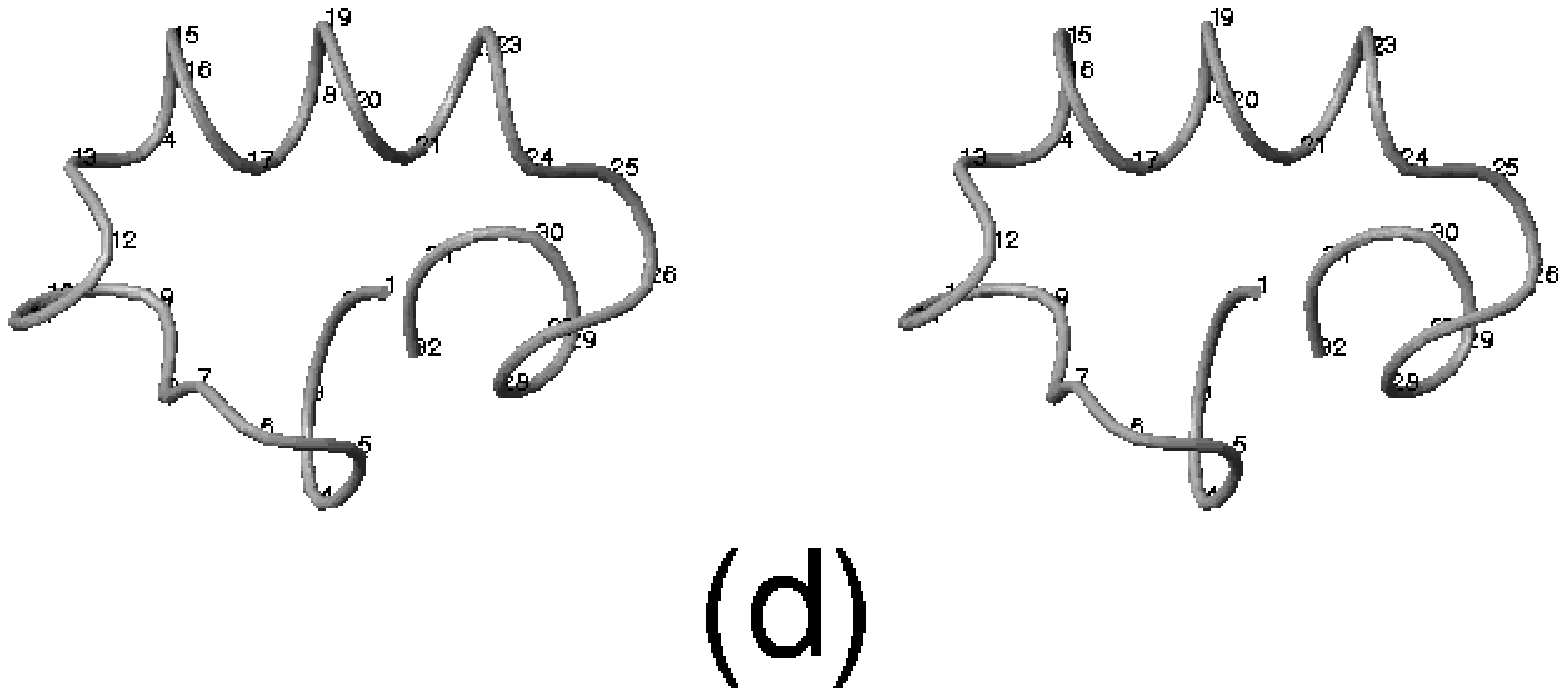}
\newpage
\epsfysize=11cm \epsfbox{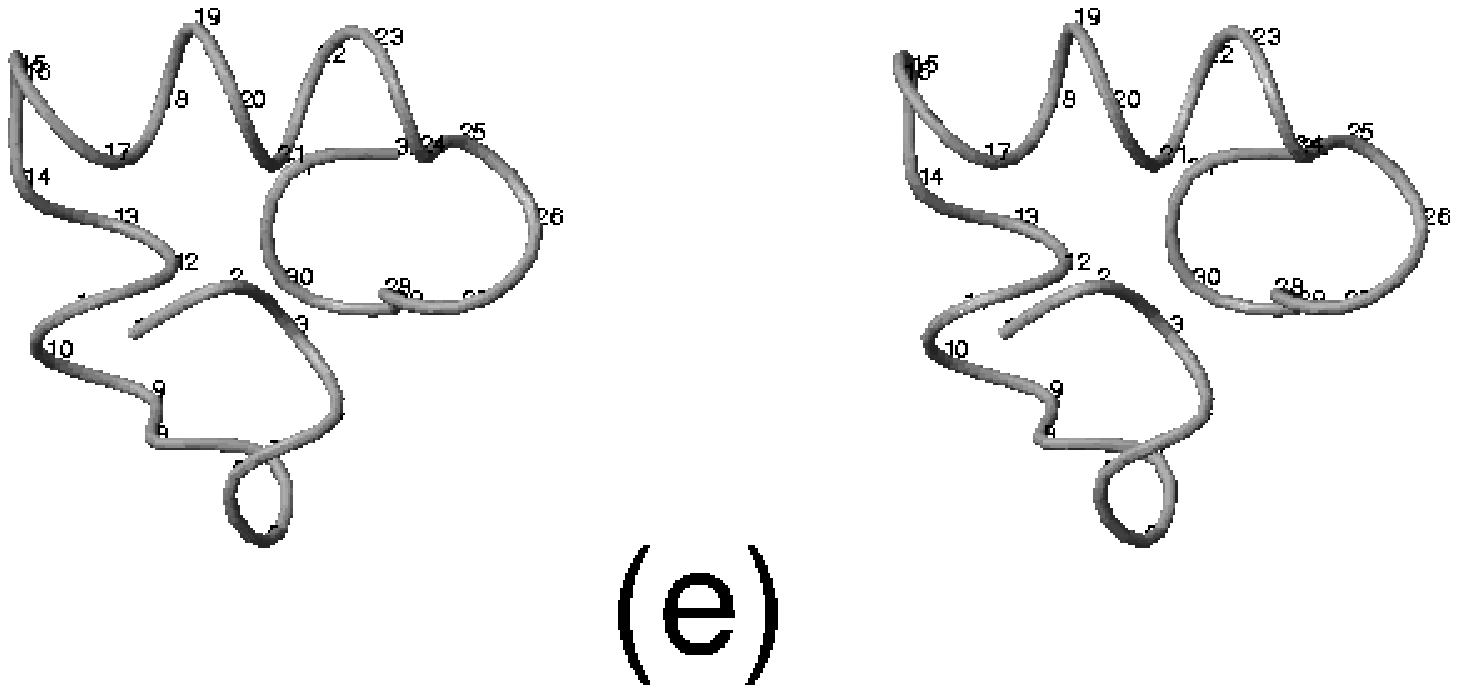}
\caption{The ${\rm C^\alpha}$
trace of the 1-32 segment of the protein 1bba. The figures were
prepared with the program MOLMOL\cite{molmol}. (a) The native
structure. The residues 15-32 form an $\a$ helix, and there is an
extended strand consisting of residues 1-12. (b) The conformation
with the lowest RMSD found with the optimized parameters. The RMSD
is $5.8$ \AA. We observe that the $\a$ helix at the C-terminal is
partially formed, consisting of residues 15-19 and 21-32. (c) The
GMEC with the optimized parameters. The RMSD is $7.9$ \AA. Again
the $\a$-helix is correctly formed except at residue 20. (d) The
lowest RMSD conformation found with the original parameters. The
RMSD is $6.2$ \AA. The position of the $\a$ helix is shifted, to
9-12, 14-24 and 27-29. (e) The GMEC with the original parameters,
with RMSD $7.6$ \AA. Again, we find $\a$ helices are formed at
wrong positions, 10-13 and 15-24. } \label{1bba}
\end{figure}

 It should be noted that a mere comparison of the RMSDs of low energy conformations found from the optimized parameters with the native structure is not so meaningful. Rather, we should check if the low energy conformation from the new parameters are closer to the native structure {\it in comparison to those from the original parameters.} We considered the 1-32 segment of the 36-residue protein 1bba. This protein contains a C-terminal $\alpha$-helix with an N-terminal extended strand parallel to the helix. The NMR structure of the protein is shown in fig.\ref{1bba}(a).
Using CSA, 200 low energy conformations are sampled for both the original parameters and the optimized parameters, respectively. The lowest RMSD values are $6.2$ \AA\ and $5.8$ \AA\ for the original and optimized parameters respectively, whereas the GMECs' RMSD values are $7.6$ \AA\ and $7.9$ \AA\ respectively. Although the RMSD values are rather large, interesting qualitative differences in the secondary structures of the sampled conformations are observed, which is difficult to be recognized by RMSD values alone. The lowest RMSD conformation and GMEC for the optimized parameters are shown in fig.\ref{1bba}(b) and (c), and those for the original parameters are shown in fig.\ref{1bba}(d) and (e). We observe that, for conformations from the optimized parameters, the $\alpha$-helices extend to the end of the C-terminal, which is in good agreement with the native structure. On the other hand, the corresponding $\alpha$-helices are incomplete near the C-terminal for the conformations obtained with the original parameters. In addition, the extended strand at the N-terminal is reproduced better with the optimized parameters.
We find these are the prevailing features of all 400 conformations we have sampled.

 We also performed Jackknife tests on other proteins, whose results are not shown here. For some $\a / \b$ proteins, notably 1L4V, the resulting conformations have similar qualitative features as above; i.e. the optimized parameters perform better in assigning secondary structures. On the other hand, the results for pure $\a$ or $\b$ proteins are not so conclusive.



\section{Conclusion and Discussion}

 We applied the general protocol for the force field parameter optimization of Lee et al.\cite{jlee}
  to the UNRES potential used in CASP3\cite{corn1,casp3}. We optimized the parameters separately for the 10-55 fragment of staphylococcal protein A (1bdd), 1fsd, and crambin (1ejg),
 and could obtain an optimal parameter set for each of them, giving the GMEC with RMSD value of $1.7$, $2.5$, $2.6$ \AA\ rrespectively.
 We could also obtain a parameter set which correctly describes the energetics of these three proteins
 simultaneously. This optimized parameter set yielded GMECs with RMSD values of $1.8$, $2.5$, and $2.6$ \AA\ for 1bdd, 1fsd, and 1ejg.

  In contrast to the earlier works\cite{jlee,pill} where
  only relative weights were optimized, we refined {\it all 709
linear parameters} of the UNRES potential.
  This enabled us to optimize the UNRES potential of Eq.(\ref{pot}) without introducing
 additional multi-body energy terms. In particular, it is demonstrated for the first time that the energetics
of proteins containing $\beta$ strands can be correctly described using the energy terms in
Eq.(\ref{pot}) only.

 It would be interesting
 to see how many proteins can be energetically well described using
  a given force field.
  This should provide a good measure for the efficacy
of existing force fields. Once the parameters for
a potential is successfully refined for the proteins in a given training set,
we should perform a Jackknife test on proteins not included in the training set.
If this test is successful, we may confidently use this potential for predicting the unknown structure of a given amino acid sequence.

 Before tackling these more challenging problems, there
 are several points in our protocol which should be improved.
 First of all, in the step of parameter refinement,
 we decreased the largest among the energy gaps of the proteins
 without any restriction, and repeated this
 procedure. However this can become quite inefficient
 as the number of proteins in the training
 set increases. We need to implement a constrained optimization where we
 require that the energy gaps of other proteins in the training set do not increase
 while that of a given protein decreases.
  Secondly, the value of RMSD cutoff at each iteration were
  determined from
  visual inspection. It would be better if one can devise a natural criterion for
  choosing RMSD cutoffs. Thirdly, in principle, one can also refine nonlinear parameters, which was not carried out in this work.
  Finally, although we considered only the UNRES potential for parameter
optimization in this work, it is straightforward to apply the
procedure to other potentials such as ECEPP\cite{ecepp4},
AMBER\cite{amber}, and CHARMM\cite{charmm} with various solvation
terms.
  All these points are left for the
  future study.

\acknowledgements{We thank Seung-Yeon Kim and Ki Hyung Joo for useful
discussions and helps in fulfilling this work. This work was carried out on our own Linux PC cluster of 76 AMD processors.}

\appendix{

\section{The UNRES potential and its linear parameters.}
The united residue(UNRES) potential is given by the
expression\cite{liwo3,liwo4,liwo5}:
\begin{eqnarray}
E&=& \sum_{i<j} U_{\rm SCSC}(i,j) + w_{\rm SCp} \sum_{i\neq j}
U_{\rm SCp}(i,j)
     + w_{\rm pp} \sum_{i<j-1} U_{\rm pp}(i,j) +
       w_{\rm b} \sum_i U_{\rm b}(i) \nonumber \\
 & & + w_{\rm tor} \sum_i U_{\rm tor}(i)+
       w_{\rm rot} \sum_i U_{\rm rot}(i)
      +w_{\rm dis} U_{\rm dis} + w^{(4)}_{\rm el-loc} \sum_{i<j} U^{(4)}_{\rm el-loc}(i,j).\label{pot2}
\end{eqnarray}

\subsection{Side-Chain Interactions}
The interactions between the side-chains are given by the
Lennard-Jones type potential: \beq U_{\rm SCSC}(i,j)=\frac{a_{\rm
SCSC}(t_i,t_j)}{r_{ij}^{12}} + \frac{b_{\rm
SCSC}(t_i,t_j)}{r_{ij}^6}, \eeq $t_i=1,\cdots 20$ being the amino
acid type of the $i$-th residue. The linear parameters we optimize
in this work are $a_{\rm SCSC}$ and $b_{\rm SCSC}$ which comprise
a total of $\frac{20\cdot 21}{2}\cdot 2=420$ parameters.

\subsection{Peptide-Peptide Interaction}
The peptide-peptide interaction is decomposed into Lennard-Jones type interaction and electrostatic interaction:
\beq
U_{\rm pp}(i,j)=U_{\rm LJ}(i,j) + U_{\rm es}(i,j)
\eeq
with the Lennard-Jones type interaction
\beq
 U_{\rm LJ}(i,j)=
 \frac{a_{\rm pp}(I_i,I_j)}{r_{ij}^{12}} + \frac{b_{\rm
pp}(I_i,I_j)}{r_{ij}^6}  \eeq and electrostatic interaction \bea
 U_{\rm es}(i,j)&=&
 \frac{a_{\rm el}(I_i,I_j)}{r_{ij}^6}[4 +(\cos\a_{ij}-3\cos\b_{ij}\cos\g_{ij})^2-3(\cos^2\b_{ij} +\cos^2\g_{ij})]\nonumber\\
&&+ \frac{b_{\rm
el}(I_i,I_j)}{r_{ij}^3}(\cos\a_{ij}-3\cos\b_{ij}\cos\g_{ij}) \eea
where $\cos\a_{ij}=(n_i \cdot n_j), \cos \b_{ij} = (n_i \cdot
r_{ji}), \cos \g_{ij} = (n_j \cdot r_{ji})$, and $n_i$ is the
vector along the $i$-th peptide. The integer $I_i$ denotes the
type of peptide group $i$. There are only two types, proline and
non-proline, with $I_i=1,2$. The term with $j=i+1$ is not included
in the summation of (\ref{pot2}) since it can be absorbed into the
local energy terms such as the bending energy. The linear parameters
we adjust for this interaction are $a_{\rm pp}$, $b_{\rm pp}$,
$a_{\rm el}$, and $b_{\rm el}$, which comprise a total of $4 \cdot
\frac{2 \cdot 3}{2}=12$ parameters.
\subsection{Side-Chain Peptide
Interaction} The side-chain peptide interaction is given by a
Lennard-Jones type interaction: \beq
 U_{\rm SCp}(i,j)=\frac{a_{\rm SCp}(I_i,t_j)}{r_{ij}^{12}} + \frac{b_{\rm SCp}(I_i,t_j)}{r_{ij}^6}
\eeq Again, $j=i\pm 1$ is not included in the summation of
(\ref{pot2}) since it can be absorbed into the local rotamer
energy. Therefore the nearest neighbors which can contribute to
this interaction are $j=i\pm2$. If we use the same parameter values for
these residues, they dominate this interaction and we get
unphysical results. Therefore one usually use smaller
parameter values for these residues in order to avoid problems. This may
seem ad-hoc, but conceptually one may justify it by noting that
for the residues  close in sequence, the quantum effect may become
important, which modifies the classical interaction parameters. Therefore, we
define additional peptide groups $I_i=3,4$ for the non-proline and
proline with $j=i\pm2$. The linear parameters to be refined are
$a_{\rm SCp}$ and $b_{\rm SCp}$, which comprise a total of $2\cdot
20 \cdot 4 = 160$ parameters.

\subsection{Disulfide Bridge Energy}
To form disulfide bridges between cysteins,  the following energy
term is introduced:
 \beq
 U_{\rm dis}= w_{\rm dis} \sum_{i'} \frac{1}{2}(D(h_1(i'),h_2(i'))-D_0)^2
\eeq
where the $i'$ and $h_1(i'),h_2(i')$ label the disulfide bridges and the residue numbers forming that bridge, respectively.
The overall weight $w_{dis}$ is the only linear parameter to be refined for this term.

\subsection{Torsional Energy}
The twist of the virtual bond between $(i-2)$-th and $(i-1)$-th
residues defines the torsion angle $\gamma_i$. Therefore the
torsion energy for $\gamma_i$ depends on the amino acid types of
these residues, which we denote by the integer $J_{i-2}$ and
$J_{i-1}$. There are three types of amino acid residues for this
interaction, glycine, proline, and the rest, with $J_i=1$, 3, 2,
respectively. For the torsion energy between two prolines, that
is, when $J_{i-2}=J_{i-1}=3$, we have \bea
U_{\rm tors}(i)&=&\sum_{j=1}^3 ( v_1(j+1,3,3)\cos(j\gamma_i)\nonumber \\
&&+v_2(j+1,3,3)\sin(j\gamma_i)\nonumber\\
&&+|v_1(j+1,3,3)| + |v_2(j+1,3,3)| \nonumber \\
&& +\cases{v_1(1,3,3)\frac{1+\cos 3\gamma_i}{1-\cos 3\gamma_i}  &{\rm for $-\frac{\pi}{3} < \gamma_i < \pi$} \cr
0 &{\rm for $-\pi \le \gamma_i \le -\frac{\pi}{3}$} \cr }
\eea
and
\bea
U_{\rm tors}(i)&=&\sum_{j=1}^6 ( v_1(j,J_{i-2},J_{i-1})\cos(j\gamma_i)\nonumber\\
&&+v_2(j,J_{i-2},J_{i-1})\sin(j\gamma_i)\nonumber\\
&&+|v_1(j,J_{i-2},J_{i-1})| + |v_2(j,J_{i-2},J_{i-1})| \eea
otherwise. The linear parameters to be refined are
$v_1(j,k,l),v_2(j,k,l)$ with $j=1$, $\cdots$, 6, $k, l=1$, 2, 3,
which comprise a total of 108 parameters.

\subsection{Local Side-Chain Energy} This energy is the negative log of a
probability distribution, which is given by the sum of Gaussian peaks:

\beq
U_{\rm rot}(i)=- w_{\rm rot} \log \Bigg[\sum_{j=1}^{n(t_i)}\exp(b_{j,t_i} \nonumber \\
-\frac{1}{2} \vec z_{j,i} G_{j,i} \vec z_{j,t_i}) \Bigg]
\eeq

where $\vec z_{j,i}\equiv \vec x_i-\vec c_{j,t_i}$, $n(t_i)$ is
the number of Gaussian peaks in the distribution which is the
function of amino acid type of the $i$-th residue $t_i$, and $\vec
x_i=(\cot \theta_i, \a_i, \b_i)$ for $k=1,2,3$, $\theta_i$,
$\a_i$, $\b_i$ being the bending angle and the polar angles of the
side-chain, respectively\cite{liwo3,liwo4,liwo5}. The values of
the non-linear parameters $b_{j,t_i}$, $c_{j,t_i}$, $G_{j,i}$ are
fixed in this work to those used in CASP3\cite{corn1,corn4,casp3},
so the only linear parameter to be refined for this term is its
overall weight $w_{\rm rot}$.

\subsection{Bending Energy} The form is similar as the local
side-chain energy, except there are two Gaussian peaks for all
amino acids types. We have

\beq
U_{\rm b}(i)=-w_b \log\Bigg[ \exp\left(-\frac{(\th_i-\th_c )^2}{2\s_c(\th_c)^2}\right)\nonumber\\
+k(\th_c)\exp\left(-\frac{(\th-\th_0(t_i))^2}{\s_0(t_i)^2}\right)\Bigg]
\eeq

where \bea
\theta_c&=&a_1(t_i)cos(\gamma_i)+a_2(t_i)sin(\gamma_i)\nonumber\\
&&+b_1(t_i)cos(\gamma_{i+1})+b_2(t_i)sin(\gamma_{i+1})\\
\sigma_c(\theta_c)^{-1}&=&2\big(p_3(t_i)\th_c^3+p_2(t_i)\th_c^2+p_1(t_i)\th_c+p_0(t_i)\big)^2\nonumber\\
&&+2 s_0(t_i)\\
k(\theta_c)&=&\exp\left(g_1(t_i)-\frac{(\th_c-g_2(t_i))^2}{2g_3(t_i)^2}\right)
\eea and again the values of the non-linear parameters $a_j(t_i)$,
$b_j(t_i)$, $p_j(t_i)$, $s_0(t_i)$, and $g_j(t_i)$ are fixed to
those used in CASP3\cite{corn1,corn4,casp3}, so the overall weight
$w_b$ is the only linear parameter to be optimized for this term.
\subsection{Multibody Term}
If $i$-th residue is in contact with $j$-th residue, this term
contributes if $(i+1)$-th residue is also in contact with
$(j+1)$-th residue or $(j-1)$-th residue. In this case, the energy
reads

\beq
U^{(4)}_{\rm el-loc}(i,j)=-p_{u(i,j),u(i+1,k)}\frac{f_{i,j}f_{i+1,k}}{r_{ij}^3 r_{i+1 k}^3}[C_+E_+(i,j) E_+(i+1,k)\nonumber\\
+ C_-E_-(j,i) E_-(i+1,k)]
\eeq

where $k=j+1$ or $k=j-1$. $f_{i,j}$
is the contact function which is $1$ when the distance between the
residues is less than a given cutoff, and $0$ when far way, and a
smooth function in the intermediate region. $C_{\pm}$ are fixed
numbers which are independent of residue types, and
$E_\pm(i,j)=E^0_+ \pm E^0_-$ where

\beq
E^0_\pm(i,j)= [4(1 \pm \cos\a_{ij}) + (\cos\a_{ij}-3\cos\b_{ij}\cos\g_{ij})^2\nonumber \\
-3(\cos\b_{ij}\pm\cos\g_{ij})^2)]^{1/2}
\eeq

and $\a_{ij},\b_{ij},\g_{ij}$ are the same as in the
peptide-peptide interaction. The integer $u(i,j)=1,2,3$ when
($i,j$) pair is (non-proline,non-proline), (non-proline,proline),
(proline,proline) respectively. We see that $p_{u(i,j),u(i+1,k)}$
comprise a total of 6 parameters since it is symmetric under the
exchange of two indices.
\subsection{Total Number of Linear
Parameters} Therefore the total number of linear parameters we
adjust is:\\
\indent
\qquad 420 {\rm (SC-SC)} + 12 {\rm (p-p)} + 160 {\rm (SC-p)} \\
\indent
\qquad + 108 {\rm (torsion)} + 6 {\rm (multi-body)}\\
\indent
\qquad + 1 {\rm (bending)} + 1 {\rm (local\ side-chain)} + 1 {\rm (disulfide\ bridge)} = 709

}

\newpage


\begin{table}
\caption{The values of RMSD cutoffs used for local CSA searches and the parameter
refinements for the case of separate optimizations (units in
\AA). The integer $i$ denotes the $i$-the iteration of CSA
search, and
$i \to i+1$ denotes the parameter refinement step
from $i$-th to $i+1$-th iteration.}
\begin{tabular}{|cccc|}
iteration & 1bdd & 1fsd & 1ejg \\
\hline
0 & (no local search)$^a$ & 3.0 & 3.0 \\
0 $\to$ 1 & 3.0 & 3.5 & 4.0 \\
1 & 2.5 & 3.0 & 2.5 \\
1 $\to$ 2 & 2.2 & 2.8, 3.0$^b$ & 3.0 \\
2 & 2.5 & 3.0 & 2.5 \\
2 $\to$ 3 & 2.2 & 2.8 & 2.5 \\
3 & 2.0 & 3.0 & 2.5 \\
3 $\to$ 4 & 2.0 & 2.7 & 2.5 \\
4 & 2.0 & 3.0 & 2.5 \\
4 $\to$ 5 & 1.8 & 2.6 &2.5 \\
5 & 2.0 & 3.0 & 2.4 \\
5 $\to$ 6 & 1.8 & 2.8 & 2.5 \\
6 & 2.0 & 3.0 & (no local search)$^a$ \\
6 $\to$ 7 & 1.8 & 2.6 & 2.5 \\
7 & 2.0 & 3.0 & (no local search)$^a$ \\
7 $\to$ 8 & 1.8 & 2.6 & 2.5 \\
8 & 2.0 & 3.0 & (no local search)$^a$ \\
8 $\to$ 9 & 1.8 & 2.6 & 2.5 \\
9 & 2.0 & 3.0 & (no local search)$^a$ \\
9 $\to$ 10 & 1.8 & 2.6 & 2.6 \\
10 & 2.0 & 3.0 & 2.5 \\
10 $\to$ 11 & 1.8 & 2.6 & 2.6 \\
11 & 2.0 & 3.0 & (no local search)$^a$ \\
11 $\to$ 12 &  & 2.6 & 2.6 \\
12 &     & 3.0 & (no local search)$^a$ \\
12 $\to$ 13 & & & 2.6 \\
13 &     & & 2.5 \\
\end{tabular}
\label{rmstable1}
$^a$ The local CSA was not carried out because the global CSA was enough to find native-like conformations.\\
$^b$ These values of the RMSD cutoff are used sequentially during the parameter refinement.
\end{table}
\begin{table}
\caption{The values of RMSD cutoff used for local CSA searches and
the parameter refinements for the case of simultaneous
optimizations (units in \AA).}
\vskip 1.0cm
\begin{tabular}{|cccc|}
iteration & 1bdd & 1fsd & 1ejg \\
\hline
0 $\to$ 1$^a$  & 1.8 & 2.6 & 2.6 \\
1 & 2.5 & 3.0 & 2.5 \\
1 $\to$ 2 & 1.8 & 2.6 & 2.6 \\
2 & 2.0 & 3.0 & 2.5 \\
2 $\to$ 3 & 1.7, 1.8$^b$ & 2.5, 2.6$^b$ & 2.55, 2.6$^b$ \\
3 & 2.0 & 3.0 & 2.5 \\
3 $\to$ 4 & 1.9, 1.8 $^b$ & 2.6, 2.5, 2.6$^b$ & 2.6, 2.5, 2.6$^b$ \\
4 & 2.0 & 3.0 & 2.5 \\
4 $\to$ 5 & 1.8 & 2.55 & 2.6 \\
5 & 2.0 & 3.0 & 2.5 \\
5 $\to$ 6 & 1.9 & 2.6 &2.6 \\
6 & 2.0 & 3.0 & 2.5 \\
6 $\to$ 7 & 1.9 & 2.6 & 2.6 \\
7 & 2.0 & 3.0 & 3.0 \\
\end{tabular}
\vskip 1.0cm $^a$ The initial conformational search is not
necessary since we use the structural databases accumulated
from the separate optimizations of three proteins.\\
$^b$ These values of the RMSD cutoff are used sequentially during the parameter refinement.
\label{rmstable2}
\end{table}


\begin{references}
\bibitem{foot2} This form of the UNRES potential was used in
CASP4\cite{corn6,casp4}.
\bibitem{foot3} To be precise, these initial conformations, which we call
first bank\cite{csa1,csa2,csa3}, are allowed to have RMSDs greater
than $R_c$. However, only new conformations with RMSDs less than
$R_c$ are allowed to update the bank.
\end{references}
\end{document}